\documentclass{iau}
\usepackage{graphicx}

\title[VGS: Galaxy Evolution and Gas Accretion in Voids] 
{The Void Galaxy Survey: Galaxy Evolution and Gas Accretion in Voids}

\author[Kreckel]   
{Kathryn Kreckel$^1$, Jacqueline H. van Gorkom$^3$, Burcu Beygu$^2$, Rien van de Weygaert$^2$, J. M. van der Hulst$^2$, Miguel A. Aragon-Calvo$^4$, Reynier F. Peletier$^2$
}

\affiliation{$^1$MPIA, K\"{o}nigstuhl 17, 69117 Heidelberg, Germany \\ email: {\tt kreckel@mpia.de} \\[\affilskip]
$^2$Kapteyn Astronomical Institute, University of Groningen, PO Box 800, 9700 AV Groningen, The Netherlands \\
$^3$Columbia University, MC 5246, 550 W120th St., New York, NY 10027, USA\\
$^4$University of California, Riverside, CA 92521, USA
}

\pubyear{2014}
\volume{308}  
\pagerange{0--0}
\jname{The Zeldovich Universe: Genesis and Growth of the Cosmic Web}
\editors{R. van de Weygaert, S. Shandarin, E. Saar \& J. Einasto, eds.}
\begin{document}

\maketitle

\begin{abstract}
Voids represent a unique environment for the study of galaxy evolution, as the lower density environment is expected to result in shorter merger histories and slower evolution of galaxies. This provides an ideal opportunity to test theories of galaxy formation and evolution. Imaging of the neutral hydrogen, central in both driving and regulating star formation, directly traces the gas reservoir and can reveal interactions and signs of cold gas accretion. For a new Void Galaxy Survey (VGS), we have carefully selected a sample of 59 galaxies that reside in the deepest underdensities of geometrically identified voids within the SDSS at distances of $\sim$100 Mpc, and pursued deep UV, optical, H$\alpha$, IR, and HI imaging to study in detail the morphology and kinematics of both the stellar and gaseous components. This sample allows us to not only examine the global statistical properties of void galaxies, but also to explore the details of the dynamical properties.  We present an overview of the VGS, and highlight key results on the HI content and individually interesting systems. In general, we find that the void galaxies are gas rich, low luminosity, blue disk galaxies, with optical and HI properties that are not unusual for their luminosity and morphology. We see evidence of both ongoing assembly, through the gas dynamics between interacting systems, and significant gas accretion, seen in extended gas disks and kinematic misalignments.  The VGS establishes a local reference sample to be used in future HI surveys (CHILES, DINGO, LADUMA) that will directly observe the HI evolution of void galaxies over cosmic time. 
\end{abstract}

\firstsection 
\section{Introduction}

The large scale clustering of galaxies into structures such as filaments, walls and clusters results in a large volume of the universe with a low density of galaxies.  These voids retain imprints from the cosmological forces which shape large scale structure, and can provide unique probes of both gravity and dark energy.  While these regions have very low density, they are not empty.  The population of void galaxies provides key constraints on our understanding of galaxy formation in a cosmological context, as well as a sample in which to study galaxy evolution in a simplified environment (see e.g. \cite[van de Weygaert 
\& Platen 2011]{vdWeygaert2011b} for a review).  Simulations only now have reached the level at which they predict not just the dark matter content of halos in a cosmological context, but the evolution of the baryonic components (stars, gas and dust) that are better suited for comparison with observations (\cite{Kreckel2011c}).

The galaxies observed to live within voids are distinct from those populating average or overdense environments.  There are very few of the most massive galaxies, most are instead lower mass, star-forming, disk galaxies (\cite[Grogin \& Geller 1999,2000; Rojas et al. 2004, 2005]{Grogin1999, Grogin2000, Rojas2004, Rojas2005}).  While they present distinct properties as a population, at fixed stellar mass they are indistinguishable from galaxies living in the `field' environment.  This suggests that in many ways galaxy evolution is largely driven by secular processes that proceed independent of the surrounding large scale environment.  This provides important insights into how galaxies evolve, and the role of external processes in driving galaxy evolution.

The lower density environment is expected to result in shorter merger histories and slower evolution of galaxies, allowing a study of how hierarchical merging and gas accretion both contribute to galaxy growth.  In particular, cold accretion of gas in the form of filamentary flows is predicted to play an important role for galaxies in low mass halos (\cite{Keres2005}).  This is expected to be most important for galaxies at high redshift, before the buildup of halos has happened, and in low density void environments at z=0, where this buildup has not yet happened.  As a result, void galaxies are the ideal targets for detailed study of the neutral gas in and around galaxies, in order to address general questions of how galaxies get their gas.  

\begin{figure}[t]
\begin{center}
 \includegraphics[width=5.3in]{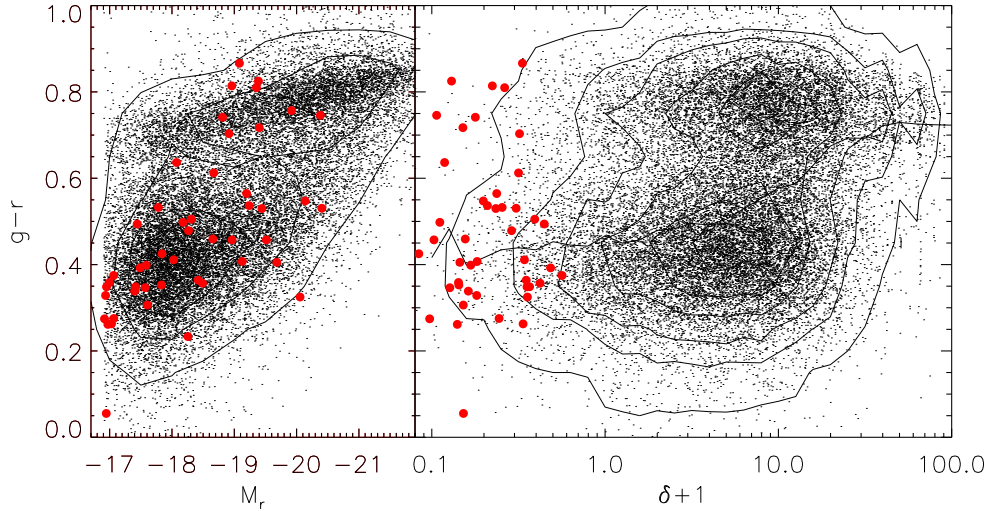} 
 \caption{Comparison of global VGS galaxy properties (color, absolute magnitude, environmental density) with a volume limited sample drawn from SDSS.  The void galaxies with redder colors are mostly edge-on and gas-rich disks, suggesting that the red colors are due to dust and not indicative of an early type morphology.  At fixed luminosity there is no evidence for bluer colors, suggesting that the color-density relation may not extend into the voids. } 
   \label{fig1}
\end{center}
\end{figure}

Dark matter simulations suggest that the voids should be threaded with low density filamentary substructure.  While these may host a substantial number of galaxies, there is some expectation that the galaxies in voids may preferentially lie within these filamentary void substructures (\cite{Sheth2004, Aragon-Calvo2013, Rieder2013}).  Tentative evidence for this has been found in the GAMA redshift survey (\cite{Alpaslan2014}), which probes to a significantly lower galaxy stellar mass range than existing redshift surveys (e.g. SDSS, 6dF).  However, the Local Void, which has been studied in great detail to very low galaxy luminosities, is remarkably empty.  Thus, the question remains whether the predicted void substructures can be traced observationally.  

We approach the study of void galaxies through the careful selection of 59 void galaxies, which we target with a full multi-wavelength series of observations.   Here we present our galaxy sample, and summarize results from our study mapping the neutral gas within and around void galaxies.   We look for evidence of ongoing gas accretion and void substructure, and find tantalizing clues within our sample.

\section{The Void Galaxy Survey}

\begin{figure}[t]
\centering
 \includegraphics[width=5.in]{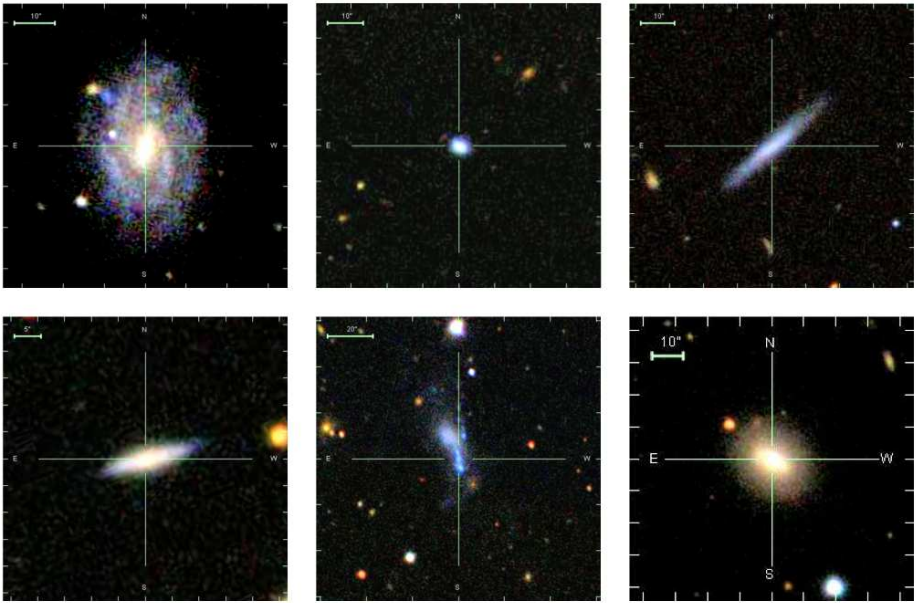} 
\caption{Sample VGS galaxies, all at the same physical scale, display a range of colors and morphologies (\cite{Kreckel2012}).}
\label{fig:samp}
\end{figure}

We identified voids within the nearby (z$<$0.025) universe using topological techniques to identify the surrounding large scale structure (\cite{Kreckel2011a}).  We apply the Delaunay Tessellation Field Estimator (\cite{Schaap2000}; \cite{vdWeygaert2009}) to the SDSS galaxy redshift survey to recover the underlying dark matter density field.  From this we use the Cosmic Spine formalism (\cite{AragonCalvo2010}) to identify walls and filaments, and the Watershed Voidfinder algorithm (\cite{Platen2007}) to identify the void boundaries.  From these, we selected 59\footnote{The sample was originally designed to be 60 galaxies, however one was later found to have its redshift mis-identified, and instead falls outside the void.} galaxies which resided in the centers of well defined voids.  Comparing the g-r color and $\delta \equiv \rho/\rho_u - 1$ density contrast of our Void Galaxy Survey (VGS) targets to a sample of SDSS galaxies (Figure \ref{fig1}, right), we see the general color-density trend, that red galaxies prefer high density environments while blue galaxies prefer low densities, and find our void galaxies well sample this lowest density population (\cite{Kreckel2012}).  

We have selected our galaxy sample independent of their intrinsic properties (e.g. color, luminosity, morphology), and find that the VGS spans a range of colors and absolute magnitudes (Figure \ref{fig1}, left).  The void galaxies with redder colors (g-r $>$ 0.6) are almost all edge-on and gas-rich disks, suggesting that the red colors are due to dust and not indicative of an early type morphology.  Upon close examination, only three of the 59 galaxies appear to have a true early type morphology. Figure \ref{fig:samp} demonstrates the range of colors and morphologies present in the sample.  The VGS does not contain any massive galaxies with absolute magnitudes brighter than M$_r$ =  -20.4 mag.  At fixed luminosity, there is no evidence for bluer colors, suggesting that the color-density relation may not extend into the voids.  This sample, however, is too small for a statistical study, and shows the need for careful controls of all factors that correlate with galaxy color.

We have compiled multi-wavelength observations for the full VGS, including GALEX NUV imaging to trace the recent (10-100 Myr) star formation, H$\alpha$ narrow band imaging to trace the current ($<$10 Myr) star formation, deep B- and R-band optical imaging to examine the outer stellar disk, Spitzer 3.6 and 4.5 $\mu$m imaging of the old stellar component (\cite{Beygu2014}),  HI line emission from the extended gas disk and 1.4 GHz radio continuum emission from star formation and AGN.  Altogether, these observations will enable us to study in detail the gas and stellar morphology and the star formation history for these galaxies to better understand their evolutionary history.  In addition, for select galaxies we have also performed near-IR and WISE 22 $\mu$m observations, tracing the old stellar population, and millimeter wavelength CO line observations, tracing the cold molecular gas that is available to form stars.

This sample is ideal for not just understanding the effect of large-scale environment on shaping these galaxies, but also, given their relatively undisturbed evolution, is well suited to study the secular processes that contribute to galaxy evolution.  

\section{Void Galaxies as Cosmological Probes}
We focus in these proceeding on results related to the neutral gas properties in the sample and what they tell us about role of gas accretion on galaxy evolution.

Our full HI survey of the VGS (\cite[Kreckel et al. 2012]{Kreckel2012}) revealed that the galaxies residing in voids typically have extended, gas-rich HI disks.  About half exhibit disturbed gas morphologies or kinematics.  However, we also find that these HI properties are typical for galaxies at the same stellar mass that are in average density environments.  This suggests that while the large scale environment is playing a role in shaping the evolution of galaxies in voids, since they tend to be low mass systems, at fixed stellar mass the galaxy HI properties are largely independent of the environment.  However, given the extremely rarified surroundings, we argue that the irregularities in the HI disks show convincing evidence for ongoing gas accretion in these systems that has proven difficult to isolate in general studies of gas in galaxies. In some cases, we see indications of gas accretion directly from out of the void.

\subsection{VGS\_12 - cold accretion of void gas?}

\begin{figure}[h]
\begin{center}
 \includegraphics[height=2.3in]{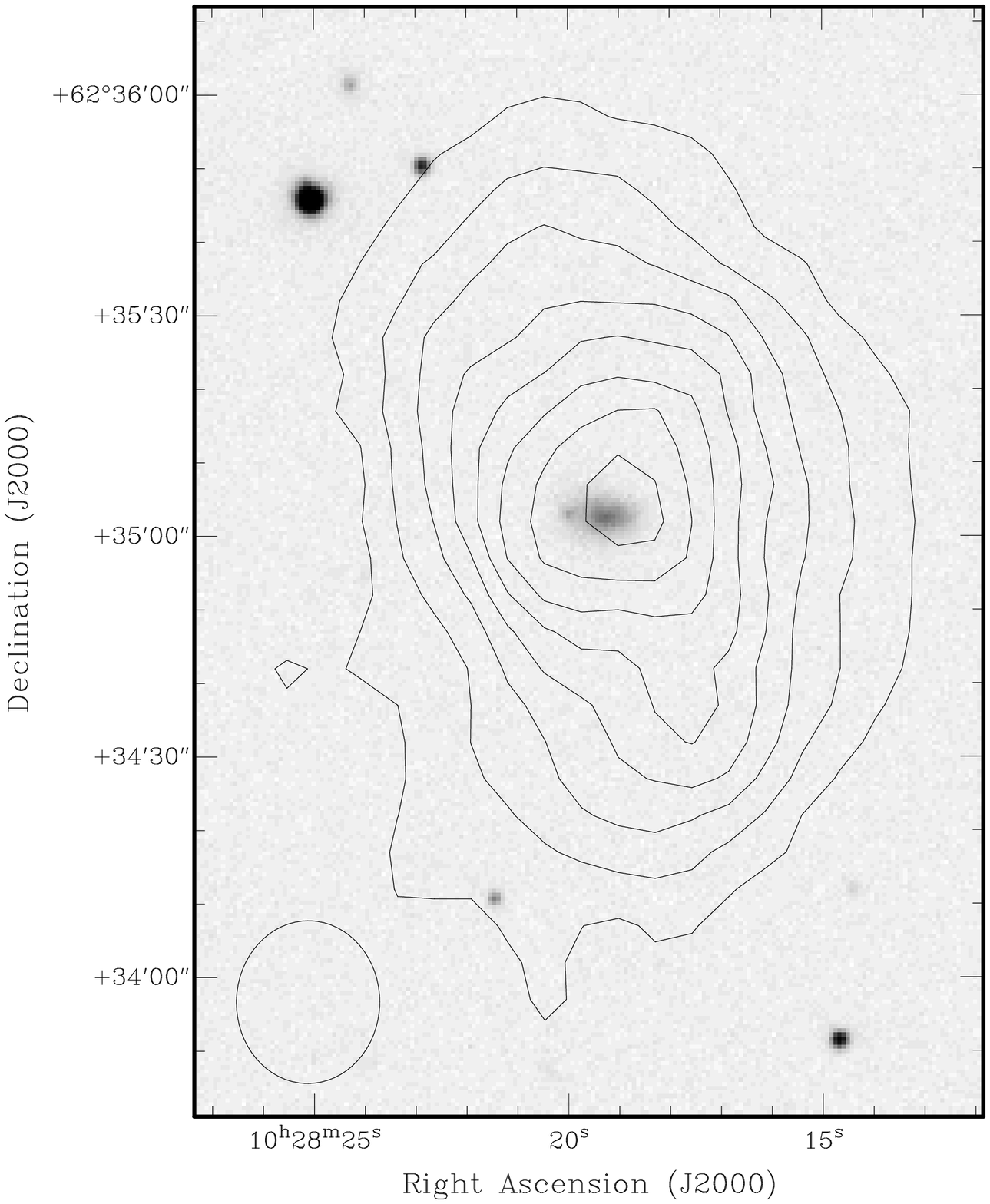} 
 \includegraphics[height=2.3in]{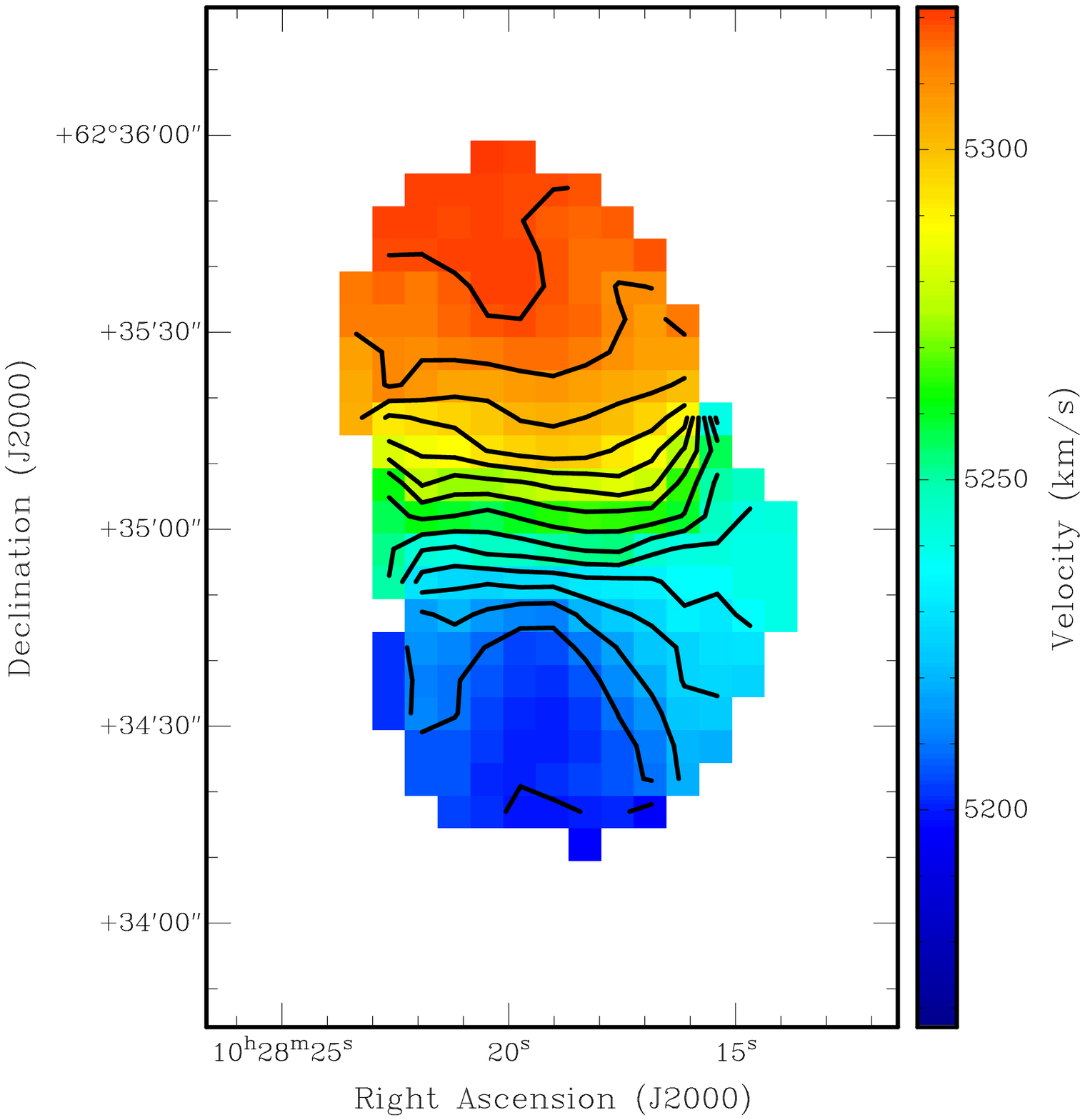} 
 \caption{VGS\_12, with HI emission shown as contours overlaid on the SDSS g-band image, and HI velocity field (Figure 1 from \cite[Stanonik et al. 2009]{Stanonik2009}, \copyright AAS. Reproduced with permission).  It exhibits an extremely extended polar disk of HI gas that is devoid of stars.  Given the pristine gas disk,  isolated nature of this system, and apparently undisturbed nature of the central stellar disk, this void galaxy presents strong evidence for ongoing cold accretion of gas onto this system.  }
   \label{fig2}
\end{center}
\end{figure}

The strongest example we have that presents evidence for accretion out of the void is VGS\_12 (Figure \ref{fig2},  \cite[Stanonik et al. 2009]{Stanonik2009}).  This small stellar system presents a blue, disk-like morphology that extends to about 5 kpc in diameter.  The HI emission also presents a regular disk-like morphology and kinematics, however it is oriented perpendicular to the stellar component, and extends over nearly 30 kpc.  This sort of polar disk configuration has been seen before in other galaxies, most notably NGC 4650A.  A satellite accretion or interaction is typically understood to be required for the formation of such disjoint kinematic components in one galaxy.  Another option is that the material is accreted directly as gas along filaments.  In all previously known cases the systems show a polar component that contains stars.  However, this is not the case for VGS\_12, where we see no evidence for a stellar component in the polar disk through deep R-band imaging.  Because of this, the isolated nature of this system, and the apparently undisturbed nature of the central stellar disk, we believe that this void galaxy presents strong evidence for ongoing cold accretion of gas onto this system.  This confirms predictions from simulations that such a mode of galaxy growth plays an important  role in low mass halos (\cite[Kere{\v s} et al. 2005]{Keres2005}), but which has proven difficult to isolate in observations.

\subsection{VGS\_31 - tracing a void filament?}

\begin{figure}[h]
\begin{center}
 \includegraphics[width=4.5in]{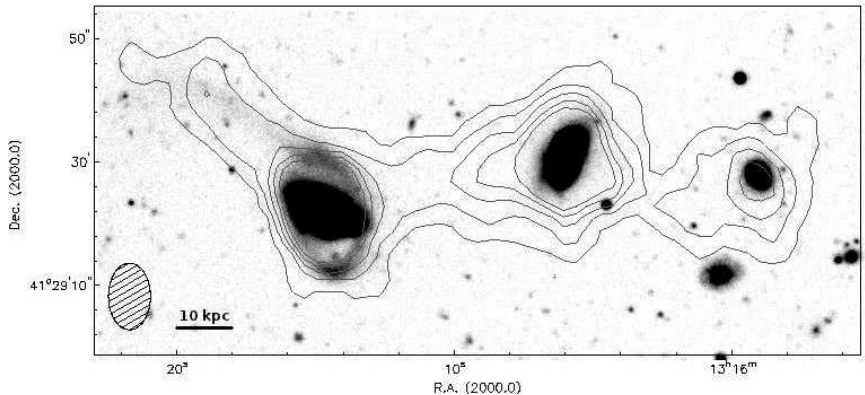} 
 \caption{The VGS\_ 31 system, with HI emission shown as contours overlaid on a deep B-band image (Figure 3 from \cite[Beygu et al. 2013]{Beygu2013}, \copyright AAS. Reproduced with permission).   It shows the possible formation of substructure within a void, as all three systems align linearly and are connected by a tenuous HI bridge. }
   \label{fig3}
\end{center}
\end{figure}

Imaging the gas morphology is crucial as it reveals features essential to understanding the formation history of these galaxies.  A second example of this is the VGS\_31 system, which contains three galaxies that appear to be forming along a filament within the void (Figure \ref{fig3}, \cite{Beygu2013}).  All three are linearly aligned and at nearly identical velocity, and they are connected by a tenuous HI bridge with relatively smooth gas kinematics connecting all three systems.  The entire filamentary configuration extends nearly 100 kpc across the sky.  Simulations designed to reproduce such a configuration show that it is likely this system is assembling within a (proto)filament and not just a group of interacting galaxies (\cite{Rieder2013}).  Further observational evidence for this is seen in the low metallicity of an outlying polar disk of material surrounding the eastern-most galaxy, suggesting again that this system presents evidence for cold accretion of gas (\cite{spavone2013}).  The VGS\_31 system presents a fascinating example of the possible formation of substructure within a void.

\subsection{KK246 - a dark void dwarf}

\begin{figure}[h]
\centering
 \includegraphics[height=2.8in]{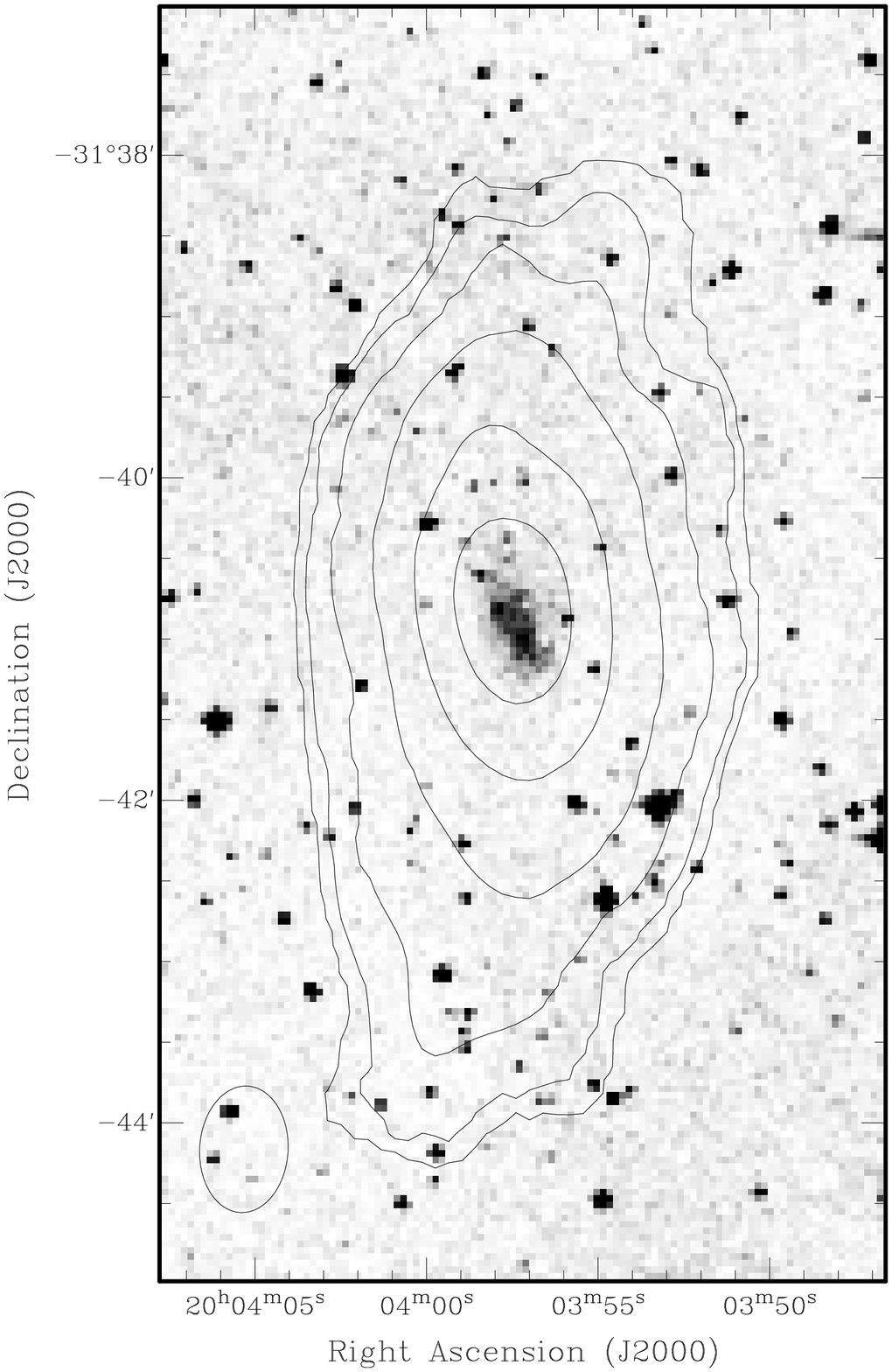} 
 \includegraphics[height=2.8in]{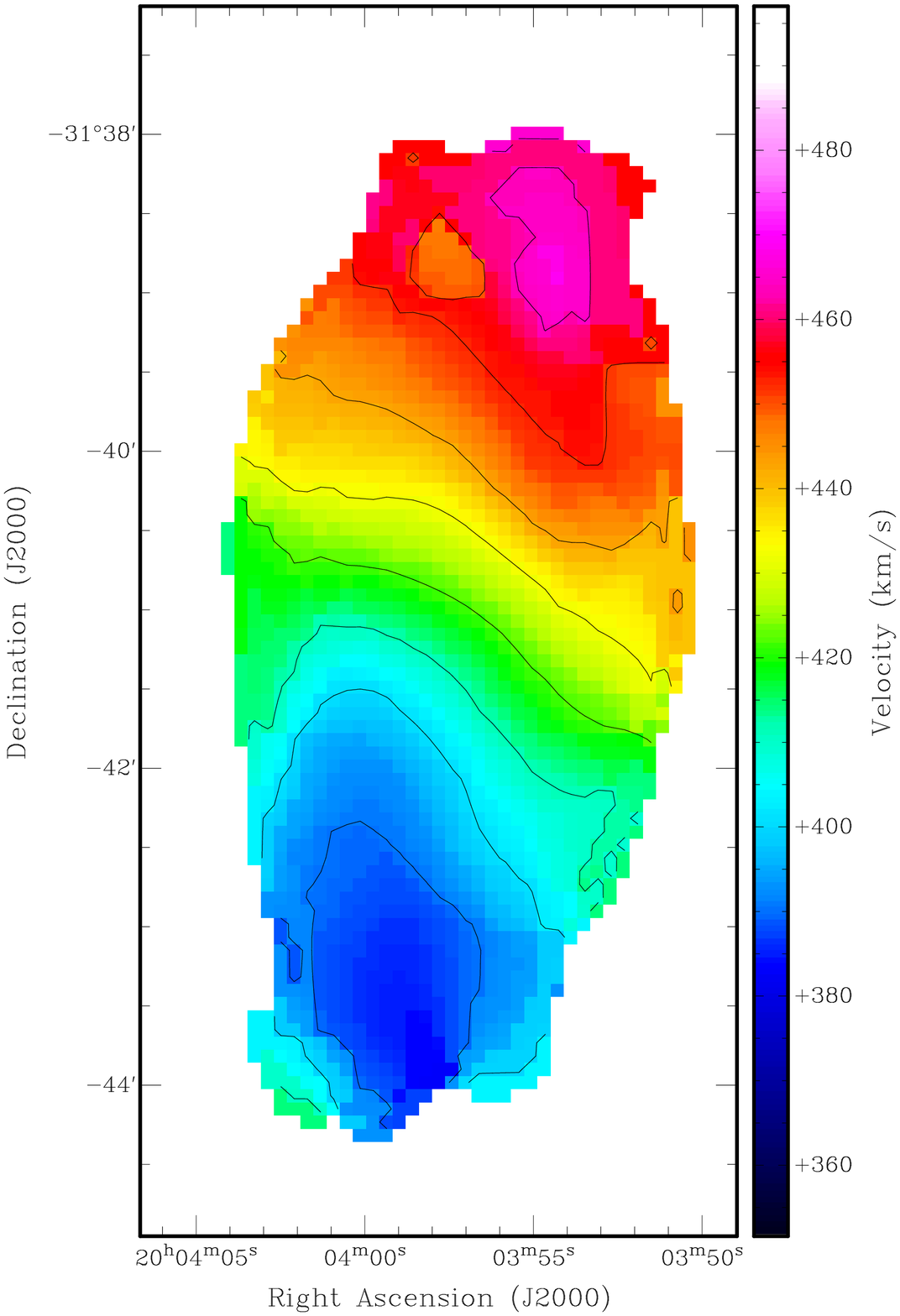} 
 \caption{KK246, with HI emission shown as contours overlaid on the B-band Palomar Observatory Sky Atlas image, and HI velocity field (Figure 1 from \cite[Kreckel et al. 2011b]{Kreckel2011b}, \copyright AAS. Reproduced with permission).  It is the only galaxy confirmed to reside well within the Local Void, and exhibits an extremely extended and misaligned gas disk.  This is suggestive of growth by accretion of gas from out of the void.  }
   \label{fig4}
\end{figure}

Void substructure is predicted to exist through the filamentary features seen threading the voids in dark matter simulations.  However, one clear counter-example is the remarkably empty Local Void (\cite{Tully2008}).  Constraints on this void are particularly strong, since its proximity allows us to identify even very faint (M$_R \sim -10$) galaxies.  HI observations of KK246, the only galaxy confirmed to reside well within this Local Void, show it also has an extremely extended and misaligned gas disk (Figure \ref{fig4}, \cite{Kreckel2011b}).   Detailed examination of the kinematics within the gas disk reveal tantalizing evidence for ongoing gas infall onto the system.  More sensitive surveys for dwarf galaxies in voids will provide important constraints on the role void substructure plays in the evolution of these systems.

Given the intriguing evidence for ongoing gas accretion in these systems, we plan to use new optical IFU observations for 10 of the more massive VGS galaxies to connect this to possible differences in the growth of the galaxy disks.  By examining the radial gradients in metallicity, stellar population and star formation history we will search for further evidence of recent pristine gas accretion. The VGS establishes a local reference sample to be used in current (CHILES, \cite{Fernandez2013}) and future HI surveys (DINGO, LADUMA) that will directly observe the HI evolution of void galaxies over cosmic time.



\begin{thebibliography}{}

\bibitem[Alpaslan et al. 2014]{Alpaslan2014} Alpaslan, M., 
Robotham, A.~S.~G., Obreschkow, D., et al.\ 2014, MNRAS, 440, L106 

\bibitem[Arag{\'o}n-Calvo et al. 2010]{AragonCalvo2010} 
Arag{\'o}n-Calvo, M.~A., Platen, E., van de Weygaert, R., 
\& Szalay, A.~S.\ 2010, ApJ, 723, 364 

\bibitem[Aragon-Calvo 
\& Szalay 2013]{Aragon-Calvo2013} Aragon-Calvo, M.~A., \& Szalay, A.~S.\ 2013, MNRAS, 428, 3409 

\bibitem[Beygu et al. 2013]{Beygu2013} Beygu, B., Kreckel, K., 
van de Weygaert, R., van der Hulst, J.~M., a
\& van Gorkom, J.~H.\ 2013, AJ, 145, 120 

\bibitem[{{Beygu}(2014)}]{Beygu2014}
{Beygu}, B. 2014, {The void galaxy survey : a study of the loneliest galaxies in the universe} ({Ph.D. Thesis, University of Groningen})

\bibitem[Fern{\'a}ndez et al. 2013]{Fernandez2013} Fern{\'a}ndez, 
X., van Gorkom, J.~H., Hess, K.~M., et al.\ 2013, ApJL, 770, L29 

\bibitem[{{Grogin} \& {Geller} 1999 }]{Grogin1999}
{Grogin}, N.~A., \& {Geller}, M.~J. 1999, AJ, 118, 2561

\bibitem[{{Grogin} \& {Geller} 2000}]{Grogin2000}
---. 2000, AJ, 119, 32

\bibitem[Kere{\v s} et al. 2005]{Keres2005} Kere{\v s}, D., 
Katz, N., Weinberg, D.~H., \& Dav{\'e}, R.\ 2005, MNRAS, 363, 2 

\bibitem[Kreckel et al. 2011a]{Kreckel2011a} Kreckel, K., Platen, 
E., Arag{\'o}n-Calvo, M.~A., et al.\ 2011a, AJ, 141, 4 

\bibitem[Kreckel et al. 2011b]{Kreckel2011b} Kreckel, K., Peebles, 
P.~J.~E., van Gorkom, J.~H., van de Weygaert, R., 
\& van der Hulst, J.~M.\ 2011b, AJ, 141, 204 

\bibitem[Kreckel et al. 2011c]{Kreckel2011c} Kreckel, K., Joung, 
M.~R., \& Cen, R.\ 2011c, ApJ, 735, 132 

\bibitem[Kreckel et al. 2012]{Kreckel2012} Kreckel, K., Platen, 
E., Arag{\'o}n-Calvo, M.~A., et al.\ 2012, AJ, 144, 16 

\bibitem[Platen et al. 2007]{Platen2007} Platen, E., van de 
Weygaert, R., \& Jones, B.~J.~T.\ 2007, MNRAS, 380, 551 

\bibitem[Rieder et al. 2013]{Rieder2013} Rieder, S., van de 
Weygaert, R., Cautun, M., Beygu, B., 
\& Portegies Zwart, S.\ 2013, MNRAS, 435, 222 

\bibitem[{{Rojas} {et~al.}(2004){Rojas}, {Vogeley}, {Hoyle}, \&
  {Brinkmann}}]{Rojas2004}
{Rojas}, R.~R., {Vogeley}, M.~S., {Hoyle}, F., \& {Brinkmann}, J. 2004, ApJ,
  617, 50

\bibitem[{Rojas et~al. 2005 {Rojas}, {Vogeley}, {Hoyle}, \&
  {Brinkmann}}]{Rojas2005}
---. 2005, ApJ, 624, 571

\bibitem[Schaap 
\& van de Weygaert 2000]{Schaap2000} Schaap, W.~E., \& van de Weygaert, R.\ 2000, A\&A, 363, L29 

\bibitem[Sheth 
\& van de Weygaert 2004]{Sheth2004} Sheth, R.~K., \& van de Weygaert, R.\ 2004, MNRAS, 350, 517 

\bibitem[Spavone \& Iodice 2013]{spavone2013} Spavone, M., \& Iodice, E.\ 2013, MNRAS, 434, 3310 

\bibitem[Stanonik et al. 2009]{Stanonik2009} Stanonik, K., Platen, 
E., Arag{\'o}n-Calvo, M.~A., et al.\ 2009, ApJL, 696, L6 

\bibitem[Tully et al. 2008]{Tully2008} Tully, R.~B., Shaya, 
E.~J., Karachentsev, I.~D., et al.\ 2008, ApJ, 676, 184 

\bibitem[van de Weygaert 
\& Schaap 2009]{vdWeygaert2009} van de Weygaert, R., \& Schaap, W.\ 2009, Data Analysis in Cosmology, 665, 291 

\bibitem[van de Weygaert 
\& Platen 2011 ]{vdWeygaert2011b} van de Weygaert, R., \& Platen, E.\ 2011, International Journal of Modern Physics Conference Series, 1, 41 

\end{thebibliography}
\end{document}